\documentclass[preprint,11pt]{elsarticle}
\usepackage{epstopdf}
\usepackage{caption}
\usepackage{setspace}
\usepackage{amssymb}
\usepackage{multirow}
\usepackage{amsthm}
\usepackage{booktabs}
\usepackage{amsmath}
\usepackage{amssymb}
\usepackage{mathrsfs}
\usepackage{longtable}
\usepackage{lscape}
\usepackage{url}
\usepackage{tabularx}
\usepackage{epsfig}
\usepackage{colortbl}
\usepackage{subfigure}
\usepackage{tikz,mathpazo}
\usepackage{graphicx}
\usepackage{natbib}
\usepackage[noend]{algpseudocode}
\usepackage{algorithm}
\usepackage{algorithmicx}   
\usepackage{array}  
\usepackage{hyperref}

\newtheorem{myDef}{Definition}

\usetikzlibrary{shapes.geometric, arrows}
\journal{International Journal of Intelligent Systems}

\begin{document}
	\begin{frontmatter}
		\title{A Fast Evidential Approach for Stock Forecasting}
		\author[address1]{Tianxiang Zhan}
		\author[address1]{Fuyuan Xiao}
		
		\address[address1]{School of Computer and Information Science, Southwest University, Chongqing, 400715, China}
		\cortext[label1]{Corresponding author: Fuyuan Xiao, School of Computer and Information Science, Southwest University, Chongqing, 400715, China. Email address: doctorxiaofy@hotmail.com.}
		\begin{abstract}
	Within the framework of evidence theory, the confidence functions of different information can be combined into a combined confidence function to solve uncertain problems. The Dempster combination rule is a classic method of fusing different information. This paper proposes a similar confidence function for the time point in the time series. The Dempster combination rule can be used to fuse the growth rate of the last time point, and finally a relatively accurate forecast data can be obtained. Stock price forecasting is a concern of economics. The stock price data is large in volume, and more accurate forecasts are required at the same time. The classic methods of time series, such as ARIMA, cannot balance forecasting efficiency and forecasting accuracy at the same time. In this paper, the fusion method of evidence theory is applied to stock price prediction. Evidence theory deals with the uncertainty of stock price prediction and improves the accuracy of prediction. At the same time, the fusion method of evidence theory has low time complexity and fast prediction processing speed.
			
		\end{abstract}
	
	\end{frontmatter}
	
	\section{Introduction}
	Stock forecasts have attracted a lot of attention recently  \cite{cao2019multi}. There are many forecasting methods used for stock forecasting, such as exponential smoothing (ES) and Holt-ES models with lower forecast accuracy. With the development of statistical science, some new forecasting methods have been developed, such as autoregressive (AR), moving forecasting (MA), moving average model (ARIMA)  \cite{tseng2002combining} and seasonal ARIMA model  \cite{tseng2002fuzzy}. These models have their limitations. In order to improve the accuracy of prediction, more and more novel and advanced methods have been developed, such as methods based on machine learning, methods based on complex networks  \cite{Zhaojy2019timeseries,Fan2019timeseries}, methods based on neural networks and so on  \cite{Xiao2021Distancemeasure,peng2020identification}.
	
	In addition, because time series are easily disturbed by external factors such as human factors or natural disasters and cannot be predicted, which will bring some errors to time series forecasting, some models based on fuzzy time series have been developed. Evidence theory is a good way to solve uncertainty, and it has received little attention in the application of time series. As a method of uncertainty reasoning, evidence theory has weaker conditions than Bayesian probability theory, and has the ability to directly express unknown knowledge  \cite{wang2018uncertainty,Deng2020InformationVolume}. Dempster–Shafer evidence theory (D-S evidence theory) also proposes the combination rule of multi-source information, namely Dempster combination rule. Dempster combination rules can integrate information provided by different time nodes to improve prediction ability  \cite{dempster2008upper,shafer1976mathematical}.
	
	This paper proposes a new method of sequence prediction inspired by evidence theory. By defining the basic belief distribution between two time points, it represents the amount of evidence provided by the current time node for future predictions. The selected time node will be used as the evidence information source to perform information fusion to obtain the amount of evidence provided by the selected time node for the future time. Finally, restore the predicted value at a future point in time. At the same time, the operating efficiency of evidence theory is much higher than traditional methods such as ARIMA  \cite{tseng2002combining}, seasonal ARIMA  \cite{tseng2002fuzzy}, complex networks, and it can be completely stable predicted method under large-scale data.
	
	The structure of this article is as follows: The second part introduces some basic concepts. The third part introduces the proposed method in detail. The fourth part explains the accuracy and performance of the proposed method. The fifth part summarizes the paper.
	
	\section{Preliminaries}
	In section, the preliminaries of evidence theory applied to time series forecasting will be introduced.
	
	As a method of uncertain modeling, evidence theory satisfies weaker conditions than Bayesian probability theory and has the ability to directly express the unknown \cite{Deng2020InformationVolume}. D-S Evidence theory also proposes a combination rule for multi-source information, the Dempster combination rule  \cite{dempster2008upper,shafer1976mathematical}. There are many directions for the expansion of evidence theory: D numbers \cite{zhang2020extension,zhou2020risk}, evidential modeling \cite{fu2020comparison,zhou2019assignment}, rule-based \cite{xu2018belief,DU20211201}, evidence theory and fuzzy set \cite{bhardwaj2021advanced,liu2021consensus,xiao2021caftr}, evidential neural network, and complex evidence theory \cite{xiao2020generalization}, and others \cite{suxiaoyan2019,Fujita2020Heuristic,meng2020uncertainty}, which has extensions in different fields \cite{deng2021fuzzymembershipfunction}. More combination methods also appeared in the later period of combination rules. Combination rules can be used for uncertain decisions makeing \cite{garg2016new,Deng2020ScienceChina}. Uncertain decisions can be made under unstable conditions in a variety of different options. Uncertain decision-making is widely used in many ways \cite{lilusu2018,Pelusi2020Improved,li2020robust,8926527}.

	Evidence theory supposes the frame of discernment($FOD$) which is the definition of a set of hypotheses as follows:
	$$\Theta =\left\{h_1,h_2,h_3,...,h_n\right\} \eqno(1)$$
	The power set $P\left(\Theta \right)$ of set $\Theta$ contains $2^n$ elements and defined as:
	$$P\left(\Theta \right)=\left\{\emptyset,\left\{h_1\right\},\left\{h_1,h_2\right\},\left\{h_1,h_2,h_3\right\},...,\Theta  \right\} \eqno(2)$$
	Basic probability assignment(BPA) is a mass function defined as $m(h)$. $m(h)$ represent the degree of the evidence support element $h$, where $h$ is an element of set $\Theta$. And the element of power set $P\left(\Theta \right)$ must satisfy the properties as follows:
	$$m\left(\emptyset\right)=0 \eqno(3)$$
	$$\sum_{X\in P\left(\Theta \right)} m\left(X\right)=1 \eqno(4)$$
	From formula $\left(4\right)$, BPA m is a mapping as follow:
	$$P\left(\Theta \right) \rightarrow \left[0,1\right] \eqno(5)$$
	
	Because BPA is effective to express uncertainty, it has been well studied, including complex mass function  \cite{Xiao2020CEQD,Xiao2021CED}, correlation coefficient  \cite{Jiang2019IJIS}, belief entropy \cite{liao2020deng}.
	
	For uncertain data, fusion is the solution to obtain more accurate information  \cite{liu2020evidence,chang2021transparent,Xiao2021GIQ,lai2020multi,liu2018classifier}. Evidence theory provides appropriate aggregation methods. For BPA $m_1,m_2,...,m_n$ obtained from multiple information sources, combination rules can be used to provide a combination of different quality knowledge  \cite{song2020selfadaptive,Fei2019Evidence}. The decision-making process can use these belief quantities and the knowledge contained in the belief function provided by each source.
	
	Dempster's combination rule is a rule defined in the framework of evidence theory, used to combine information from different sources  \cite{dempster2008upper,shafer1976mathematical}. The combination result of BPAs is represented by $m$, and the definition of combination BPA as follows:
	$$m\left(A\right)=\frac{\sum \prod^{i=n}_{X_1\cap X_2\cap ...\cap X_n=A} m_i\left(X_i\right) }{1-k} \eqno(6)$$
	$$m\left(\emptyset\right)=0 \eqno(7)$$
	$$k=\sum \prod^{i=n}_{X_1\cap X_2\cap ...\cap X_n=\emptyset} m_i\left(X_i\right) \eqno(8)$$
	which $n$ is size of set of sources' elements and $1-k$ can be simplified as follow:
	$$1-k=1-\sum \prod^{i=n}_{X_1\cap X_2\cap ...\cap X_n=\emptyset} m_i\left(X_i\right)$$
	
	$$=\sum \prod^{i=n}_{X_1\cap X_2\cap ...\cap X_n\neq \emptyset} m_i\left(X_i\right) \eqno(9)$$
	
	Dempster's combination rule  \cite{dempster2008upper,shafer1976mathematical} was useful for uncertainty processing and applied in many areas, like fault diagnosis  \cite{Xiao2020maximum}, data fusion  \cite{Xiao2020evidencecombination}.
	
	\section{The proposed method}
	In this part, the new time series forecasting method is presented and it includes the two steps:
	
	Step 1 (Convert time series to BPA model) For discrete time series $U=\left \{ \left ( t_1,y_1 \right ),\left ( t_2,y_2 \right ),...,\left ( t_n,y_n \right ) \right \}$, each time point $T_n=\left ( t_n,y_n \right )$ in $U$ can be converted to a BPA. $m_n\left(A\right)$ represent the degree of the evidence support $T_{n+1}$ , and each $T_n$ provides evidence for the $T_{n+1}$.
	\begin{myDef}
		The similar confidence function $m_m\left(A\right)$  is defined as follow:
		$$m_n\left(A\right)=\frac{y_{n}}{y_{n+1}} \eqno(10)$$
		$$m_n\left(\bar{A}\right)=1-m_n\left(A\right)=1-\frac{y_{n}}{y_{n+1}} \eqno(11)$$
		which each $T$ source set $S$ as follow:
		$$S=\left\{\emptyset,A,\bar{A},\left\{A,\bar{A}\right\}\right\} \eqno(12)$$
		Here, the amount of similar confidence function provided by the empty set and the complete set of absolute conflict is 0.
		$$m\left(\emptyset\right)=m\left(\left\{A,\bar{A}\right\}\right)=0 \eqno(13)$$
	\end{myDef}
	 
	If $y_m>y_{m+1}$, then $m_m\left(A\right)>1$ and $m_m\left(\bar{A}\right)<0$, that means the time point $T_m$ provided the excessive evidence to predict the time point $T_{m+1}$. Also, $y_m=y_{m+1}$ and $y_m<y_{m+1}$ is represented that the time point $T_m$ provided the equal evidence to predict the time point $T_{m+1}$ and the time point $T_m$ provided the short evidence to predict the time point $T_{m+1}$.  BPA reflects the amount of evidence provided by the previous node to the next node. The evidence is the value at the previous point in time. In the  evidence theory, the BPA can not over 1, but in the time series,  the evidence of the previous time point can be more than the time point needed now which means that n time prediction, BPA can exceed 1.
	
	Step2 (Data fusion of the time series) Time series $U$ has $n$ values, and there are a BPA set $B$ with $\left(n-1\right)$ BPA sources which converted from each $\left(T_{n},T_{n+1}\right)$ in the Time series $U$ as follow:
	$$B=\left\{m_1,m_2,...,m_{n-1}\right\} \eqno(14)$$
	
	A single BPA represent the degree of the evidence support next time point. Therefore, for a whole time series, the degree of the evidence support next time point is a combination of each point ahead of last time point.
	\begin{myDef}
		The combined BPA from set $B$ represent by $\hat{m}$ called "$\ Global\ $
		\\$belief\ assignment\left(GBPA\right)$" as follows:
		$$m_{GBPA}\left(U\right)=\frac{\sum \prod^{n}_{y_1\cap y_2\cap ...\cap y_n=A} m_i\left(y_i\right) }{1-k} \eqno(15)$$
		
		$k$ is the conflict between the evidences, called conflict probability.
		$$k=\sum \prod^{n}_{y_1\cap y_2\cap ...\cap y_n=\emptyset} m_i\left(y_i\right) \eqno(16)$$
		$$1-k=1-\sum \prod^{n}_{y_1\cap y_2\cap ...\cap y_n=\emptyset} m_i\left(y_i\right)$$
		$$=\sum \prod^{i=n}_{y_1\cap y_2\cap ...\cap y_n\neq \emptyset} m_i\left(y_i\right) \eqno(17)$$
	\end{myDef}
	When calculating $1-k_{new}$, each time a new time point is added, the result of the previous $k_{old}$ calculation process can be saved:
		$$1-k_{new}=\sum \prod^{i=n}_{y_1\cap y_2\cap ...\cap y_n\neq \emptyset} m_i\left(y_i\right) = \sum (1-k_{old})_i*m_i\left(y_i\right)\eqno(18)$$
		
	$(1-k_{old})_i$ is the value before the accumulation of each previous intersection as an empty set, and $m\left(y_i\right)$ is the confidence function that the intersection is an empty set after adding a point. Also, $\sum \prod^{n}_{y_1\cap y_2\cap ...\cap y_n=A}m_i\left(y_i\right)$ can be calculated in the same way. Through this mechanism, the proposed algorithm has the asymptotic time complexity $O\left(1\right)$ in continuous time series prediction.

	According to definition of GBPA, there must be a value of evidence source combined all sources from time series $U$ called "$ global\ value\left(GV\right)$". The GV is not only a combination of each point value  $y$ in each time point $T$ but a combination of time value $t$ in the same point $T$.
	\begin{myDef}
		For a time point $T_i$, the evidence called "$Evidential\ value$$\left(EV\right)$" is defined as follows:
		$$y^{EV}_{i}=y_n+\frac{y_n-y_i}{t_n-t_i}(t_{n+1}-t_n) \eqno(19)$$
	\end{myDef}
	
	Then by aggregating the evidence values of each time point into a EV set $E$, set $E$ is as follows:
	$$E=\left\{y^{EV}_{1},y^{EV}_{2},...,y^{EV}_{n-1}\right\} \eqno(20)$$
	
	\begin{myDef}
		GV is defined as follow:
		$$y^{GV}=\frac{\sum_{y^{EV}\in E} y^{EV}}{n-1} \eqno(21)$$
	\end{myDef}
	
	So the GV is combined both point values and time values. The predicted value can be recoveried by GBPA and GV as follow which is a inverse operation of formula 10:
	$$\hat{y}_{n+1}= y^{GV} \ast m_{GBPA}\left(U\right) \eqno(22)$$
	
	$\hat{y}_{n+1}$ is the prediction value of time series $U$. Here is a simple example:
	
	Suppose the time series U is $U=\left\{10,12,11,14,10,15\right\}$. Then calculate the BPA and EV at each time point, and the results are as follows:
	
	\begin{table}[htbp]
	\centering
	\setlength{\tabcolsep}{1mm}
		\begin{tabular}{ccccccc}
			\hline
			Index      & 1     & 2     & 3     & 4     & 5     & 6                \\ \hline
			Data value & 10.00 & 12.00 & 11.00 & 14.00 & 10.00 & 15.00            \\
			$m(A)$       & 0.83  & 1.09  & 0.79  & 1.40  & 0.67  & \textbackslash{} \\
			$m(\bar{A})$       & 0.17  & -0.09 & 0.21  & -0.40 & 0.33  & \textbackslash{} \\
			EV       & 15.67  & 15.8 & 15.73  & 15.93 & 15.67  & \textbackslash{} \\ \hline
		\end{tabular}
		\caption{BPA of example time series}
	\end{table}

	The GBPA and GV of the current time series can be calculated from the calculation results in the above table.
	$$y^{GV}=\frac{\sum_{y^{EV}\in E} y^{EV}}{5}=15.76 \eqno(23)$$
	$$m_{GBPA}\left(U\right)=\frac{\prod m_i\left(A\right)}{1-k}=\frac{0.6667}{0.6671}=0.9994 \eqno(24)$$
	
	The calculation method of $1-k$ is as follows:
	$$1-k=(1-k)_1+(1-k)_2=0.6671 \eqno(25)$$
	$$(1-k)_1=\prod m_i\left(A\right)=0.6667 \eqno(26)$$
	$$(1-k)_2=\prod m_i\left(\bar{A}\right)=0.0004 \eqno(27)$$
	
	The predicted value of this time series can be calculated through the GV and GBPA of the time series:
	$$\hat{y}_{n+1}= y^{GV} \ast m_{GBPA}\left(U\right)=15.75 \eqno(28)$$
	
	If this time series data is updated, the seventh time node $U_7=16$ is added. Firstly, the BPA and EV of the newly added data needs to be updated.
	\begin{table}[htbp]
		\centering
		\setlength{\tabcolsep}{1mm}
		\begin{tabular}{cccccccc}
			\hline
			& 1       & 2       & 3       & 4       & 5       & 6       & 7                \\ \hline
			Data value & 10      & 12      & 11      & 14      & 10      & 15      & 16               \\
			$m(A)$       & 0.8333  & 1.0909  & 0.7857  & 1.4000  & 0.6667  & 0.9375  & \textbackslash{} \\
			$m(\bar{A})$       & 0.1667  & -0.0909 & 0.2143  & -0.4000 & 0.3333  & 0.0625  & \textbackslash{} \\
			EV         & 16.6250 & 16.7500 & 16.6875 & 16.8750 & 16.6250 & 16.9375 & \textbackslash{} \\ \hline
		\end{tabular}
	\caption{Updated BPA of example time series}
	\end{table}

	Then update the GV and GBPA of the time series U:
	$$y^{GV}=\frac{\sum_{y^{EV}\in E} y^{EV}}{6}=16.75 \eqno(29)$$
	$$m_{GBPA}\left(U\right)=\frac{\prod m_i\left(A\right)}{1-k}=\frac{0.6250}{0.625025}=0.99996 \eqno(30)$$
	
	The updated method of $1-k$ is as follows:
	$$1-k=(1-k)_1+(1-k)_2=0.625025 \eqno(31)$$
	$$(1-k)_1=(1-k_{old})_1*m_6(A)=0.6667*0.9375=0.6250 \eqno(32)$$
	$$(1-k)_2=(1-k_{old})_2*m_6(\bar{A})=0.0004*0.0625=2.5*10^{-5} \eqno(33)$$
	
	The predicted value of the updated time series can be calculated through the updated GV and GBPA of the time series:
	$$\hat{y}_{n+1}= y^{GV} \ast m_{GBPA}\left(U\right)=16.75 \eqno(34)$$

	The algorithm and flowchart of the proposed method are in Fig.1 below.
	
	\begin{algorithm}[htbp]
		\caption{The Process of the Proposed Method When Running for the First Time }
		\label{alg::conjugateGradient}
		\begin{algorithmic}[1]
			\Require
			Time series dataset $T$ :$ N$ data values;
			\Ensure
			The prediction value of $\hat{y}_{N+1}$
			\For {time point T= 1 to n-1}
			\State Calculate $m(A)$;
			\State Calculate $m(\bar{A})$
			\State Calculate $y^{EV}_i$;
			\EndFor
			\State Use $y^{EV}_i$, calculate the $y^{GV}$
			\State Data fusion of $m(A)$, calculate the $m_{GBPA}$	
			\State Calculate the $\hat{y_{n+1}}$
			\Return $\hat{y}_{n+1}$	
		\end{algorithmic}
	\end{algorithm}

	\begin{algorithm}[htbp]
		\caption{The Process of the Proposed Method When Updating Data }
		\label{alg::conjugateGradient}
		\begin{algorithmic}[1]
			\Require
			Time series dataset $T$ :$ N'$ data values;
			\Ensure
			The prediction value of $\hat{y}_{N+1}$
			\State Calculate $m(A)$ of the new data;
			\State Calculate $m(\bar{A})$ of the new data;
			\State Calculate $y^{EV}$ of the new data;
			\State Use $y^{EV}_i$, calculate the $y^{GV}$
			\State Use $(1-k_{old})i$, calculate the $1-k$
			\State Calculate the $m_{GBPA}$	
			\State Calculate the $\hat{y_{n+1}}$
			\Return $\hat{y}_{n+1}$	
		\end{algorithmic}
	\end{algorithm}

	\begin{figure}[htbp]
		\centerline{\includegraphics[scale=0.8]{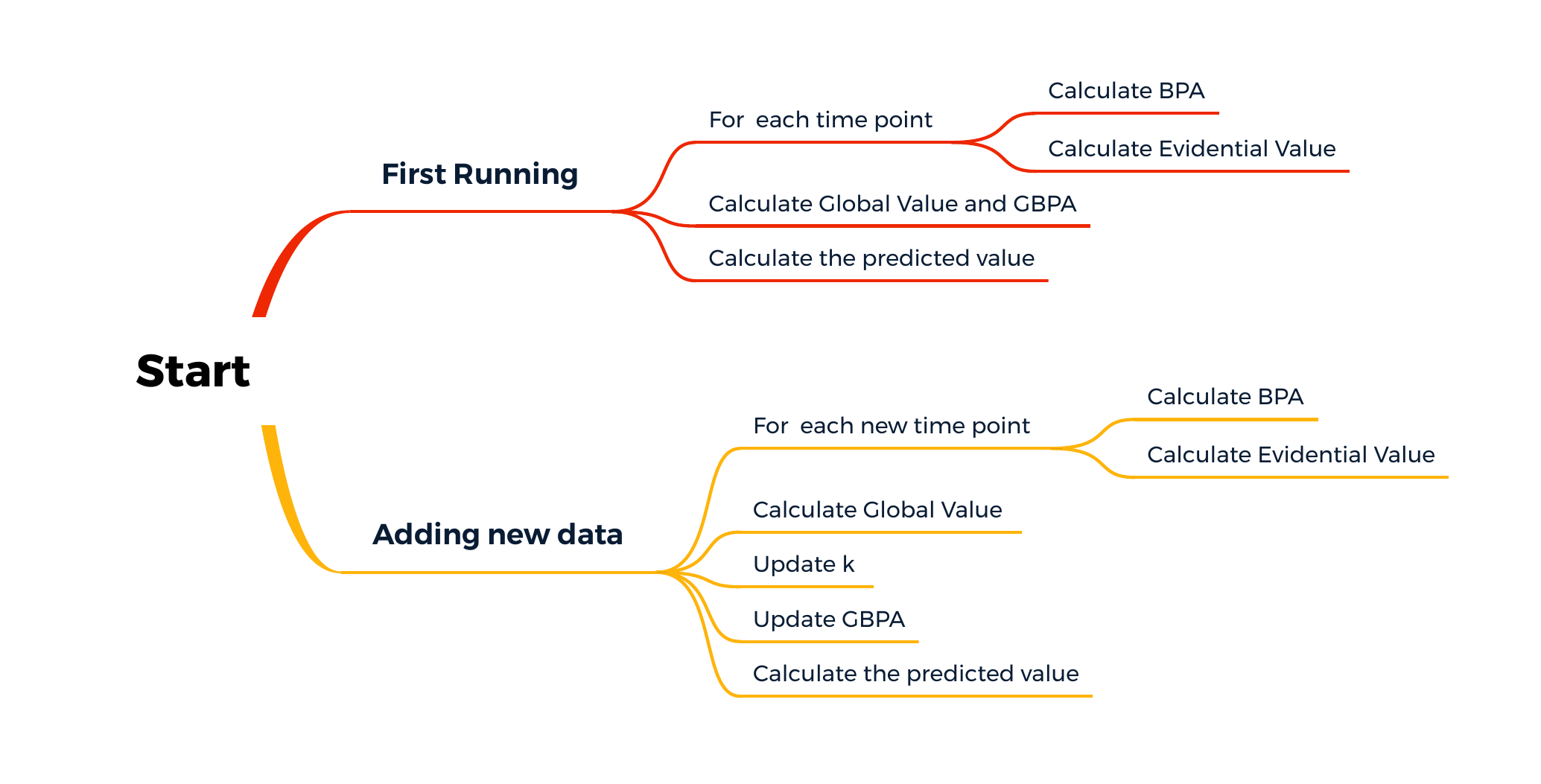}}
		\caption{The flowchart of the proposed method}
	\end{figure}
		
	\section{Experiments}
	In this section, the proposed method in this article will be applied to time series come from economic area. At the same time, this method also has good forecasting effects in other fields, such as epidemic forecasting and index forecasting. To judge the prediction of each size, there are five measures of error: mean absolute difference (MAD), mean absolute percentage error (MAPE), root mean square error (RMSE), and normalized root mean squared error (NRMSE):
	$$MAD=\frac{1}{N}\sum_{t=1}^{N}\left|\hat{y}(t)-y(t)\right| \eqno(35)$$
	$$MAPE=\frac{1}{N}\sum_{t=1}^{N}\frac{\left|\hat{y}(t)-y(t)\right|}{y(t)} \eqno(36)$$
7	$$RMSE=\sqrt{\frac{1}{N}\sum_{t=1}^{N}\left|\hat{y}(t)-y(t)\right|^{2}} \eqno(37)$$
	$$NRMSE=\frac{\sqrt{\frac{1}{N}\sum_{t=1}^{N}\left|\hat{y}(t)-y(t)\right|^{2}}}{y_{max}-y_{min}} \eqno(38)$$
	$$SMAPE=\frac{2}{N}\sum_{t=1}^{N}\frac{\left|\hat{y}(t)-y(t)\right|}{\hat{y}(t)+y(t)} \eqno(39)$$
	where $\hat y(t)$ is the predicted value, $y(t)$ is the true value and N is the total number of $\hat y(t)$. The experiments are conducted  on a Lenovo Xiaoxin Pro-13-2020 personal laptop with 16GB RAM @3200 MHz and an AMD Ryzen 7 4800U with Radeon Graphics CPU @1.80 GHz multi-core processor under Windows 10.

	In order to prove that the prediction of this method is more accurate, the comparison method chooses the method of complex network(Zhang et al. method and Mao and Xiao's method)  \cite{zhang2017novel,mao2019time} ,simple moving average (SMA)  \cite{guan2017two}, ARIMA  \cite{tseng2002combining} and seasonal ARIMA  \cite{tseng2002fuzzy}. The first 4 time points of the experimental time series are used as the basic time series.
	
	\subsection{TAIEX forecasting}
	In this experiment, the Taiwan Capital Appreciation Stock Index (TAIEX) will be used as a data set in the investment field \cite{yu2005weighted}. TAIEX often undergoes great changes, attracting many investors and economists. The TAIEX data set used in this experiment is from January 5, 1967 to June 3, 2017, with a total of 13,778 pieces of TAIEX data. TAIEX is a relatively large data set, which can reflect the superiority of the proposed method.  Fig.2 is an intuitive diagram of predicted data and actual TAIEX data. By comparing the predicted data and the real TAIEX data, the trends of the two curves are basically the same. The prediction curve is very close to the real curve. Under the condition of large data volume and frequent jitter, the prediction is quite accurate.

\begin{figure}
	\centerline{\includegraphics[scale=1]{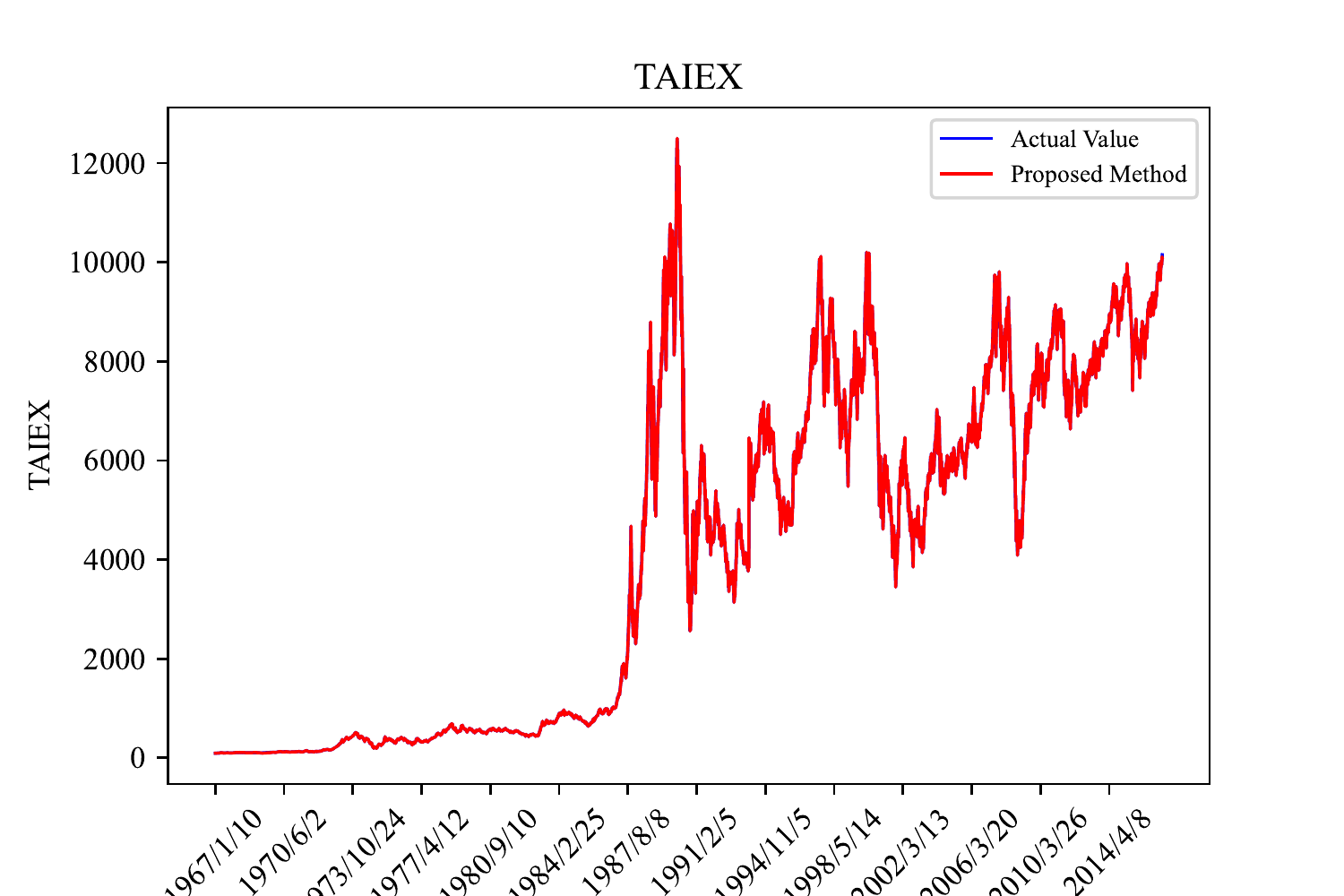}} 
	\centerline{\includegraphics[scale=1]{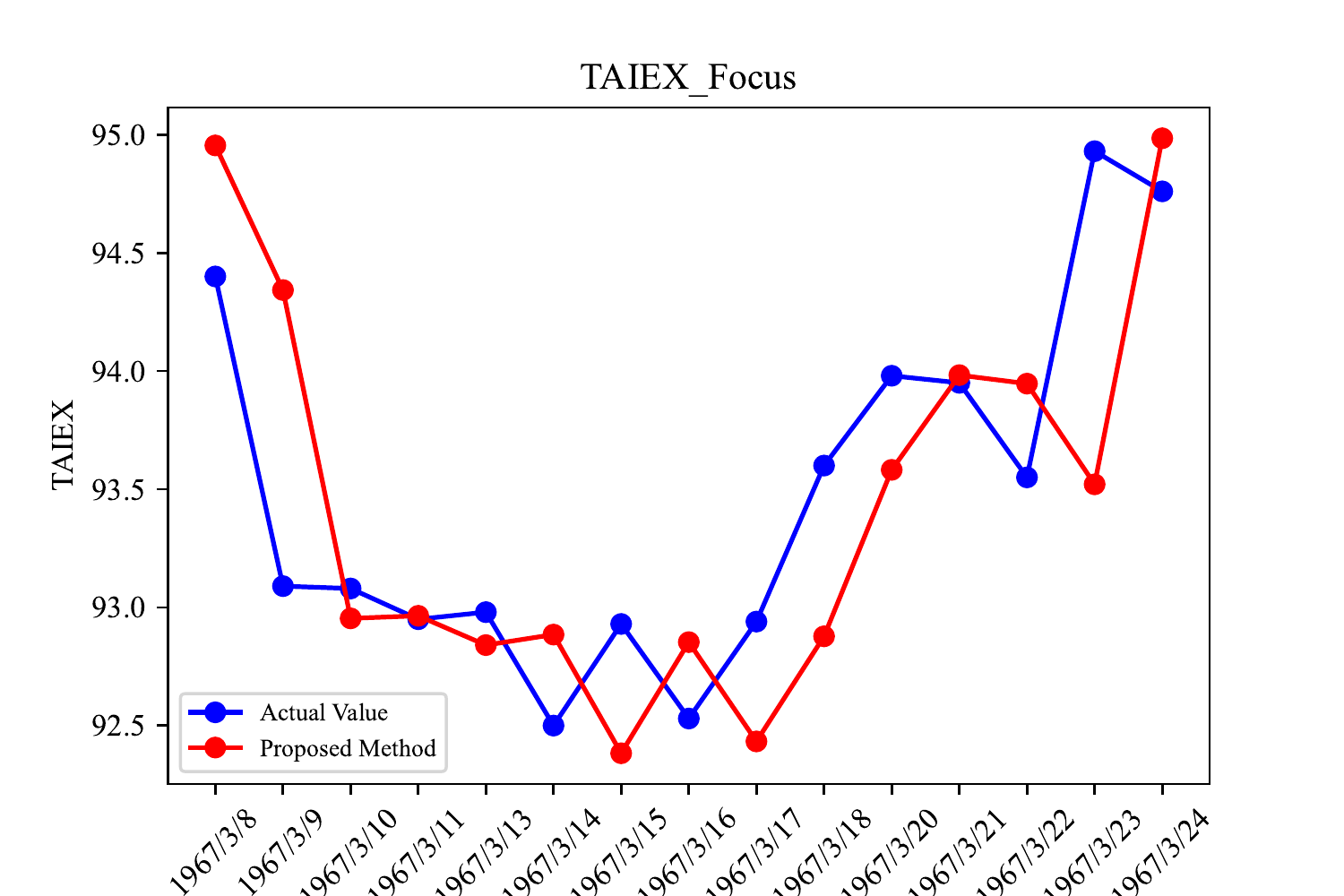}}
	\caption{The TAIEX and predicted values} 
\end{figure}

\begin{figure}

	\centerline{\includegraphics[scale=1]{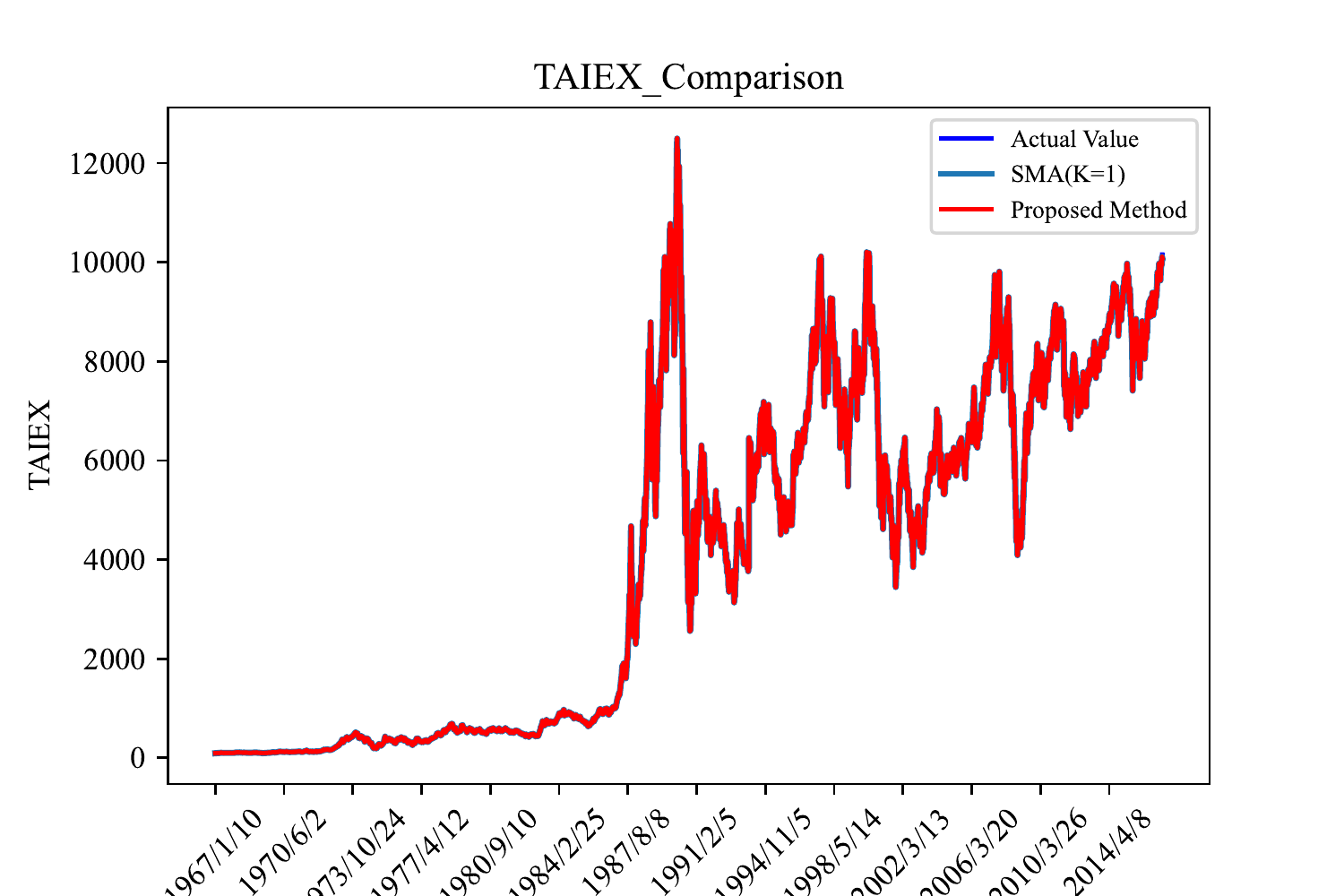}}
	\centerline{\includegraphics[scale=1]{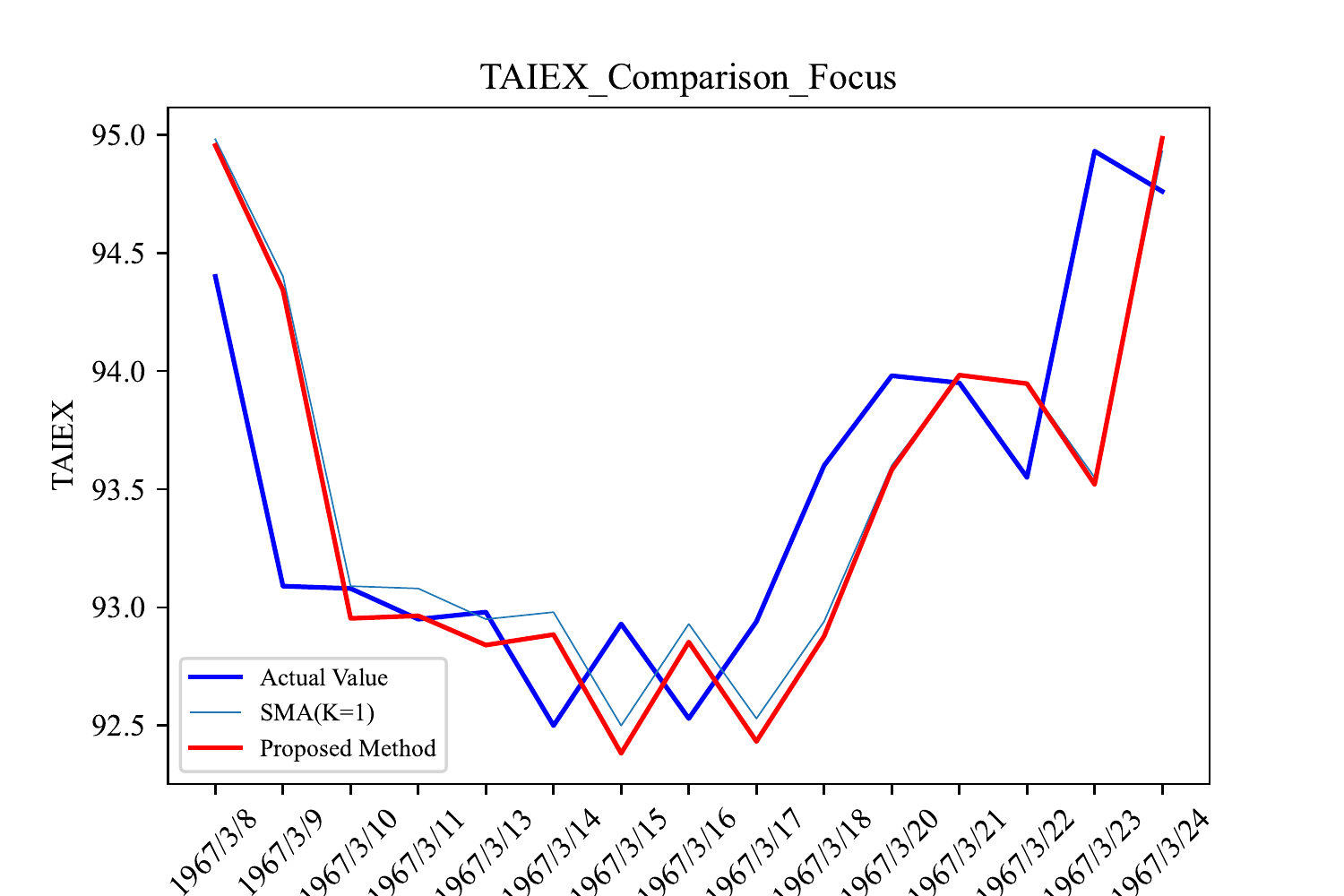}}
	\caption{The TAIEX and comparison of predicted values} 
\end{figure}
	
	For the big TAIEX data prediction, the complex network method cannot be carried out due to the high complexity of the algorithm and ARIMA  \cite{tseng2002combining} and seasonal ARIMA  \cite{tseng2002fuzzy} need to reselect parameters according to the updated time series so they also cannot be carried out. The proposed method and the prediction error of SMA  \cite{guan2017two} are described in Tab.1 below:

	\begin{table}[htbp]
	\centering
	\setlength{\tabcolsep}{1mm}
	\begin{tabular}{cccccc}
			\hline
			Predication methods      & MAD              & RMSE    & MAPE(\%)        & NRMSE(\%)       & SMAPE(\%)       \\ \hline
			SMA(K=1)  \cite{guan2017two}                & 44.2823          & 82.9004 & 1.0335          & 0.6684          & 1.0338          \\
			Zhang et al.   \cite{zhang2017novel}            & -                & -       & -               & -               & -               \\
			Mao and Xiao  \cite{mao2019time}            & -                & -       & -               & -               & -               \\
			ARIMA    \cite{tseng2002combining}          & -                & -       & -               & -               & -               \\
			Seasonal ARIMA    \cite{tseng2002fuzzy}            & -                & -       & -               & -               & -               \\
			\textbf{Proposed method} & \textbf{44.2532} & 82.9011 & \textbf{1.0326} & \textbf{0.6684} & \textbf{1.0328} \\ \hline
		\end{tabular}
		\caption{The errors of the proposed method in the prediction of TAIEX}
	\end{table}

	The proposed method can fuse trend information at various time points for time series forecasting. TAIEX is a kind of ordinary jitter data. The trend of data change is large and frequent. The data predicted by SMA \cite{guan2017two} only predicts the future time point through one time point, without considering the characteristics of the data itself. The proposed method can combine the historical trend to synthesize relatively accurate prediction results by updating the trend at a new time point, but both prediction and data fusion will cause errors.

	\subsection{Asymptotic time complexity analysis}
	In time series forecasting, different forecasting methods have different forecasting performance. In today's huge amount of stock data calculation, operating performance is a very important reference variable. The TAIEX data set has a total of 13,778 pieces of data. Compared with the general small data set, many prediction methods are difficult to calculate the expected results normally under the condition of continuous prediction. Although some forecasting methods have high forecasting accuracy, the algorithms are too complex to be competent for large-scale continuous data forecasting. The number of consecutive predictions of TAIEX is 13,773. The following table shows the asymptotic time complexity and actual running time of SMA  \cite{guan2017two}, Zhang et al.  \cite{zhang2017novel}, Mao and Xiao  \cite{mao2019time} and the proposed method:
	
	\begin{table}[htbp]
	\centering
	\setlength{\tabcolsep}{1mm}
		\begin{tabular}{ccc}
			\hline
			Method                   & Asymptotic time complexity & Actual Time                    \\ \hline
			SMA(K=1)   \cite{guan2017two}               & $O\left(1\right)$          & 0.01second          \\
			Zhang et al.  \cite{zhang2017novel}            & $O\left(n^3\right)$        & \textgreater{}1day             \\
			Mao and Xiao  \cite{mao2019time}            & $O\left(n^3\right)$        & \textgreater{}1 day            \\
			ARIMA  \cite{tseng2002combining}            & $O\left(n\right)$        & \textgreater{}1 day            \\
			Seasonal ARIMA   \cite{tseng2002fuzzy}           & $O\left(n\right)$        & \textgreater{}1 day            \\
			\textbf{Proposed method} & \textbf{$O\left(1\right)$} & \textbf{0.03second} \\ \hline
		\end{tabular}
		\caption{The asymptotic time complexity of the proposed method and the actual running time}
	\end{table}

	The asymptotic time complexity of the proposed method is far less than that of the complex network method. The complex network method will produce time series to complex network conversion and random walk process, so the time complexity is high and the running performance is low. In continuous forecasting, whenever a new time point is added to the time series, the proposed method can inherit the previous calculation result and continue to use, so the time complexity is low, and it has excellent forecasting performance and forecasting effect.
	
	The D-S combination rule provides a fusion method of multiple information. The proposed method treats each time point as an information source. Combining the proposed similar confidence function, the uncertainty of multiple time points can be integrated to obtain a relatively accurate The trend of growth or decline. Methods such as SMA and ARIMA are only pure function fitting and do not consider the relationship between time points. The method based on the complex network will lead to low prediction efficiency due to the data structure of the network. At the same time, the method of the complex network only considers the connectivity at each time. Compared with the fusion method of evidence theory, the prediction accuracy is poor.

	\subsection{Effects in other areas}
	\subsubsection{CCI forecasting}
	The Construction Cost Record (ENR) publishes the Construction Cost Index (CCI) once a month \cite{hwang2011time}. The CCI data for the construction industry is worthy of reference data and has been studied by many scholars in the construction industry. In this experiment, a total of 295 CCI data values (the CCI data set from January 1990 to July 2014) are used to predict the time series. Fig.4 is an intuitive diagram of predicted data and actual CCI data. Under the conditions of numerical jitter, the fit of the method can be represented in the Fig.4. In the prediction of the CCI, the predicted data do not differ much from the actual CCI data. 
		
	\begin{figure}
		\centerline{\includegraphics[scale=1]{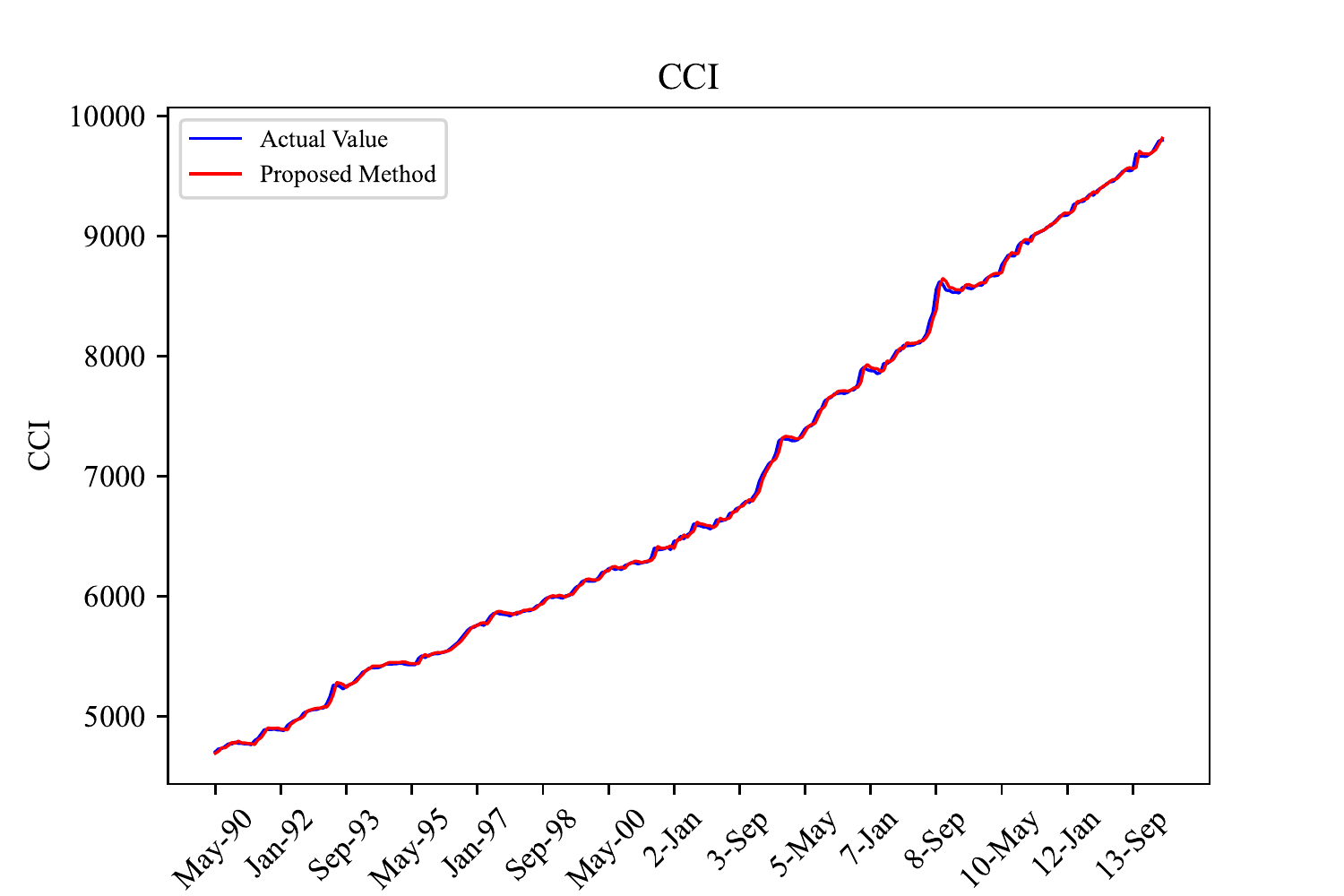}} 
		\centerline{\includegraphics[scale=1]{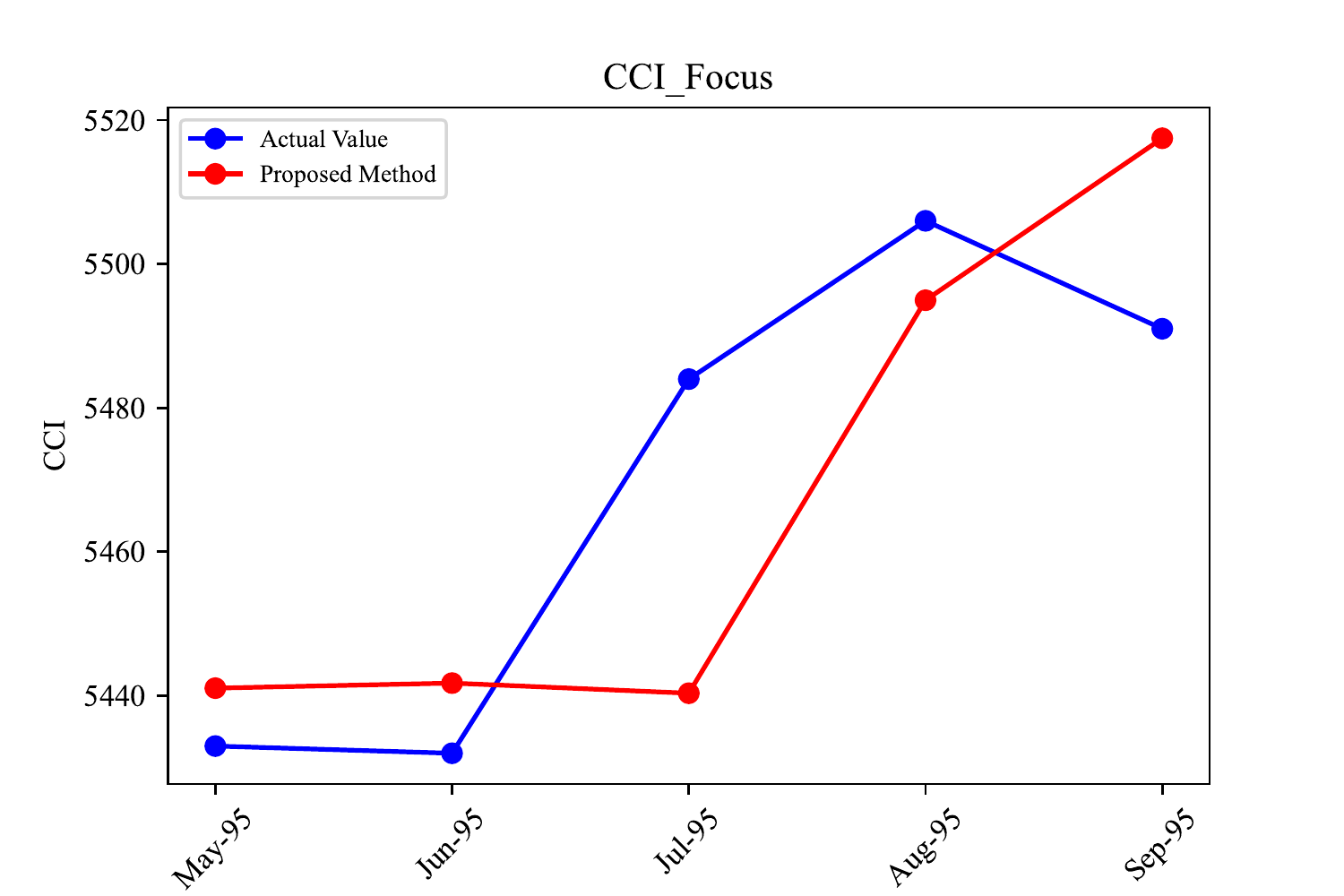}}
		\caption{The CCI and predicted values} 
	\end{figure}
	
	\begin{figure}
		
		\centerline{\includegraphics[scale=1]{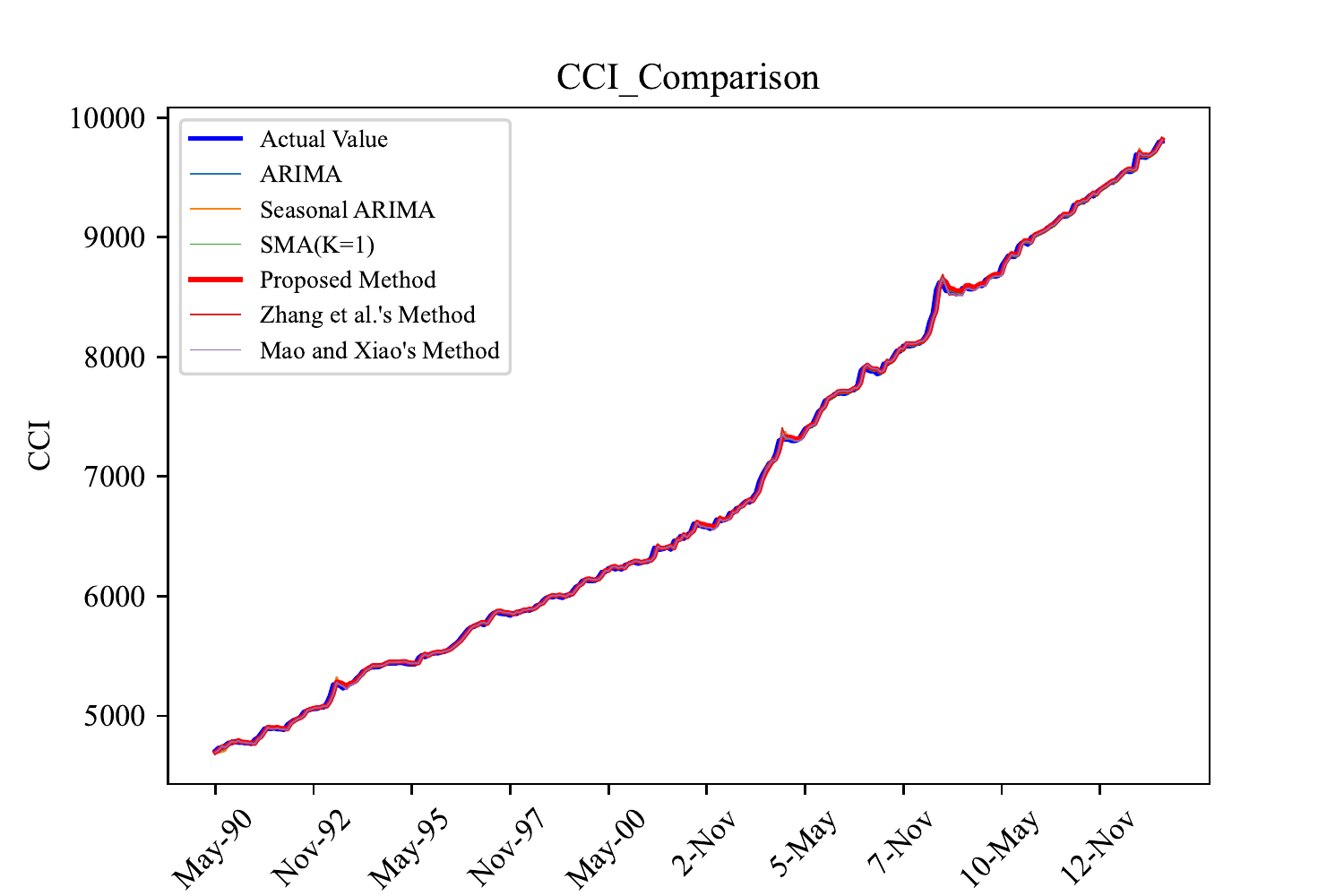}}
		\centerline{\includegraphics[scale=1]{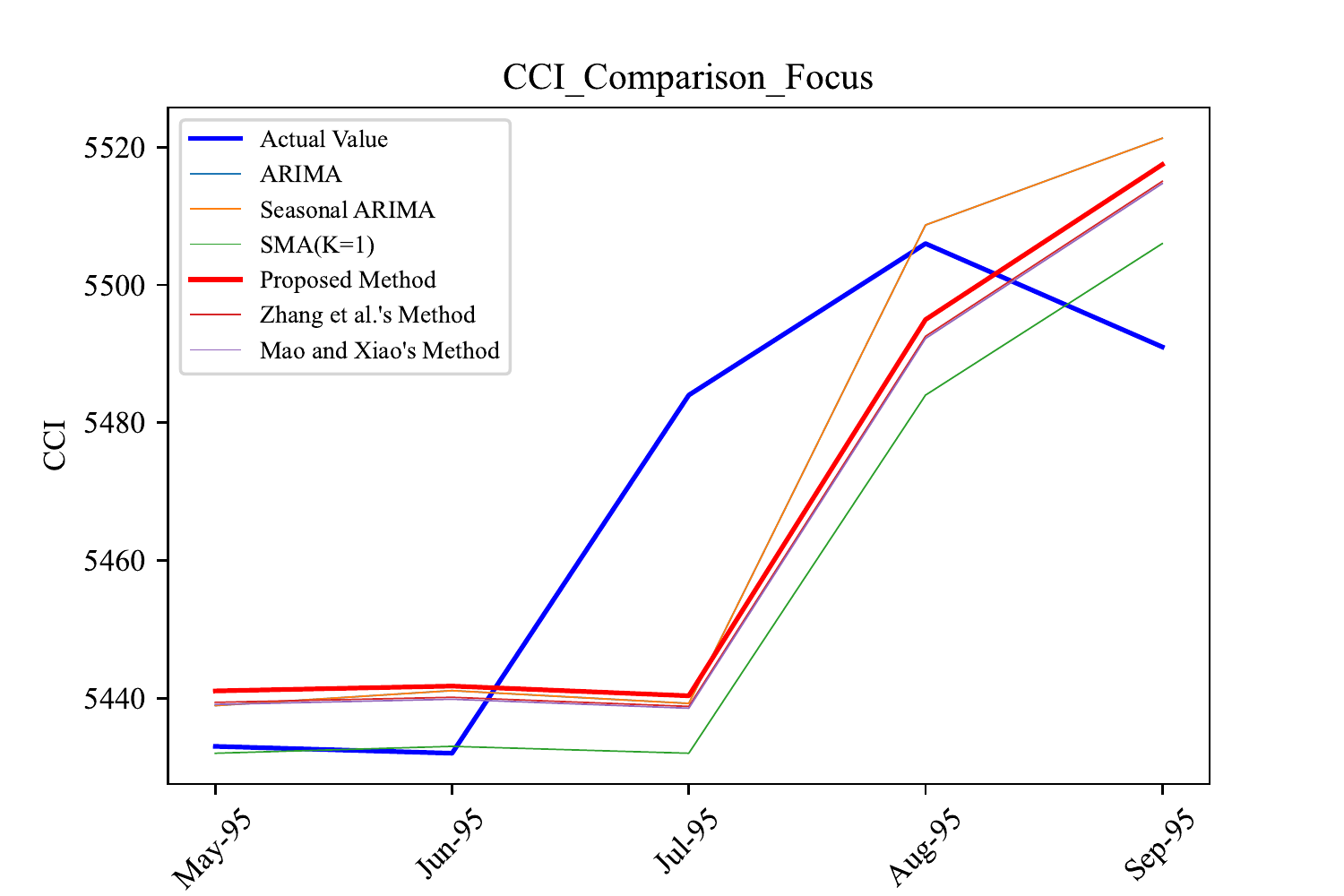}}
		\caption{The CCI and comparison of predicted values} 
	\end{figure}
	
	On the CCI data set, all three prediction methods can predict normally. The forecast error is shown in the table below:

	\begin{table}[htbp]
			\centering
		\setlength{\tabcolsep}{1mm}
		\begin{tabular}{cccccc}
			\hline
			Prediction Method        & MAD     & MAPE   & SMAPE  & RMSE             & NRMSE            \\ \hline
			SMA(k=1)  \cite{guan2017two}                 & 21.6529 & 0.3122 & 0.3132 & 32.8336          & 45.9447          \\
			Zhang et al.  \cite{zhang2017novel}            & 19.7469 & 0.2860 & 0.2865 & 28.7903          & 40.2868          \\
			Mao and Xiao   \cite{mao2019time}           & 19.3613 & 0.2807 & 0.2812 & 28.2348          & 39.5095          \\
						ARIMA  \cite{tseng2002combining}                    & 20.5531 & 0.3006 & 0.3009 & 28.8274          & 40.3387          \\
			Seasonal ARIMA  \cite{tseng2002fuzzy}           & 20.5309 & 0.3004 & 0.3007 & 28.7544          & 40.2366          \\
			\textbf{Proposed Method} & 19.6787 & 0.2855 & 0.2859 & \textbf{27.7506} & \textbf{38.8320} \\ \hline
		\end{tabular}
	\caption{The errors of the proposed method in the prediction of CCI}
	\end{table}
	
	\subsubsection{Enrollment forecasting}
	In this experiment, the number of students at the University of Alabama was used as a small amount of data collection as the experimental object of the proposed method \cite{chen1996forecasting}. As a general data collection, the registry data from 1971 to 1992 was selected, and the experiment completed the data prediction. Fig.5 is an intuitive diagram of predicted data and actual data. It can be found that as the sampling points increase, the fitting curve is getting closer and closer to the true value. The actual number of registrations and the number of expected registrations of this method have experienced almost the same trend of change. Tab.4 lists the experimental error of the method and the experimental error of the comparison method.
		
	\begin{figure}
		\centerline{\includegraphics[scale=1]{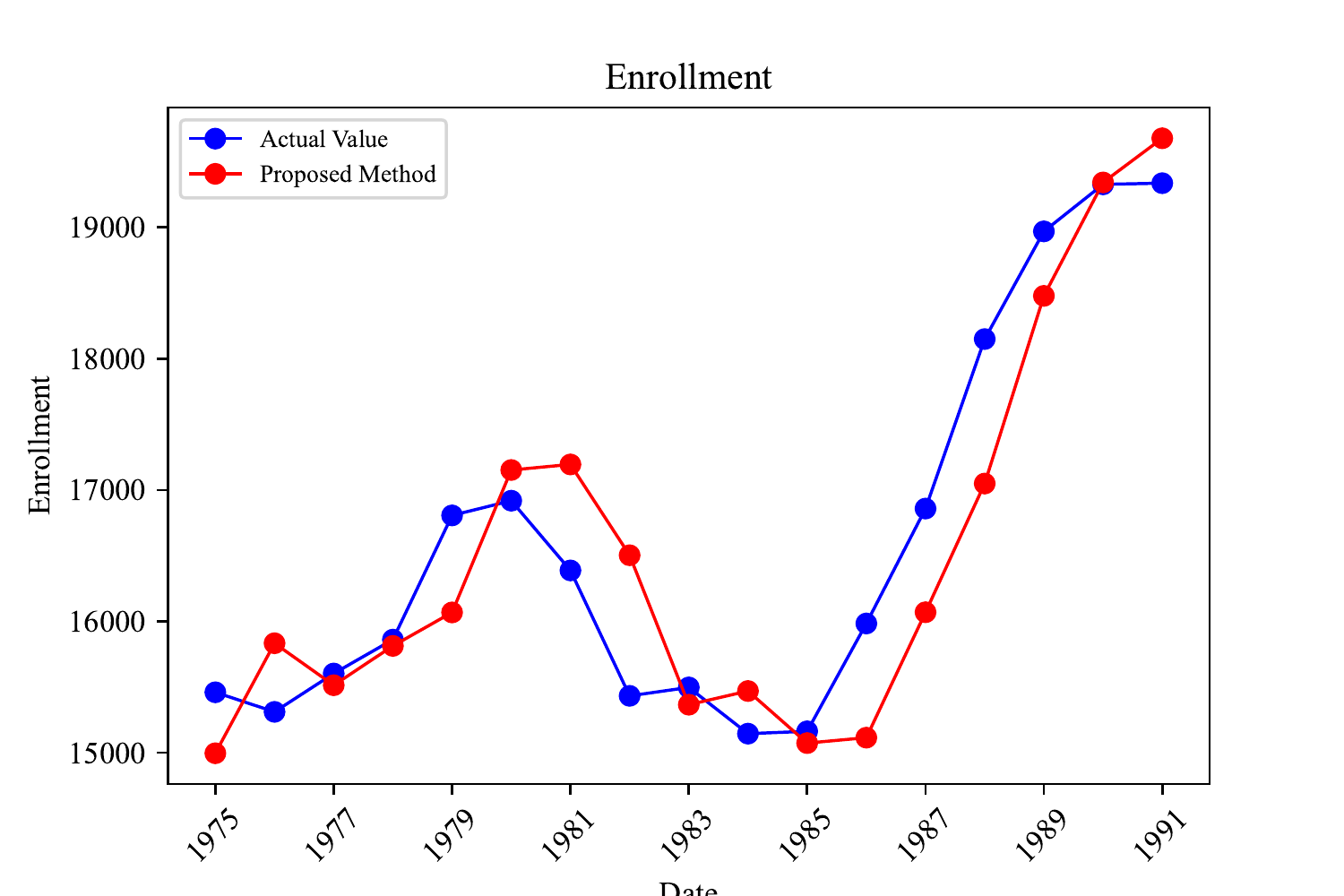}} 
		\centerline{\includegraphics[scale=1]{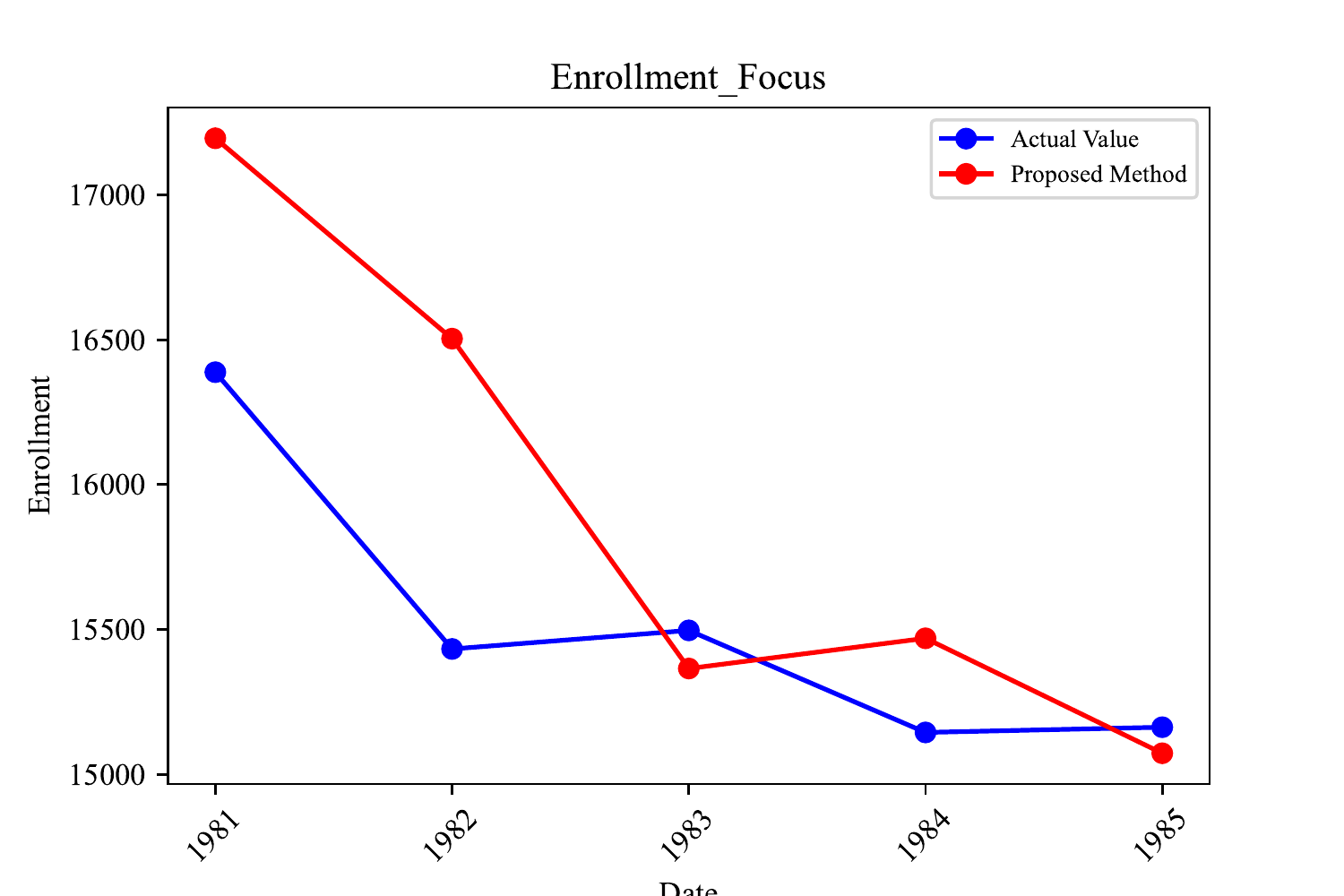}}
		\caption{The Enrollment and predicted values} 
	\end{figure}
	
	\begin{figure}
		
		\centerline{\includegraphics[scale=1]{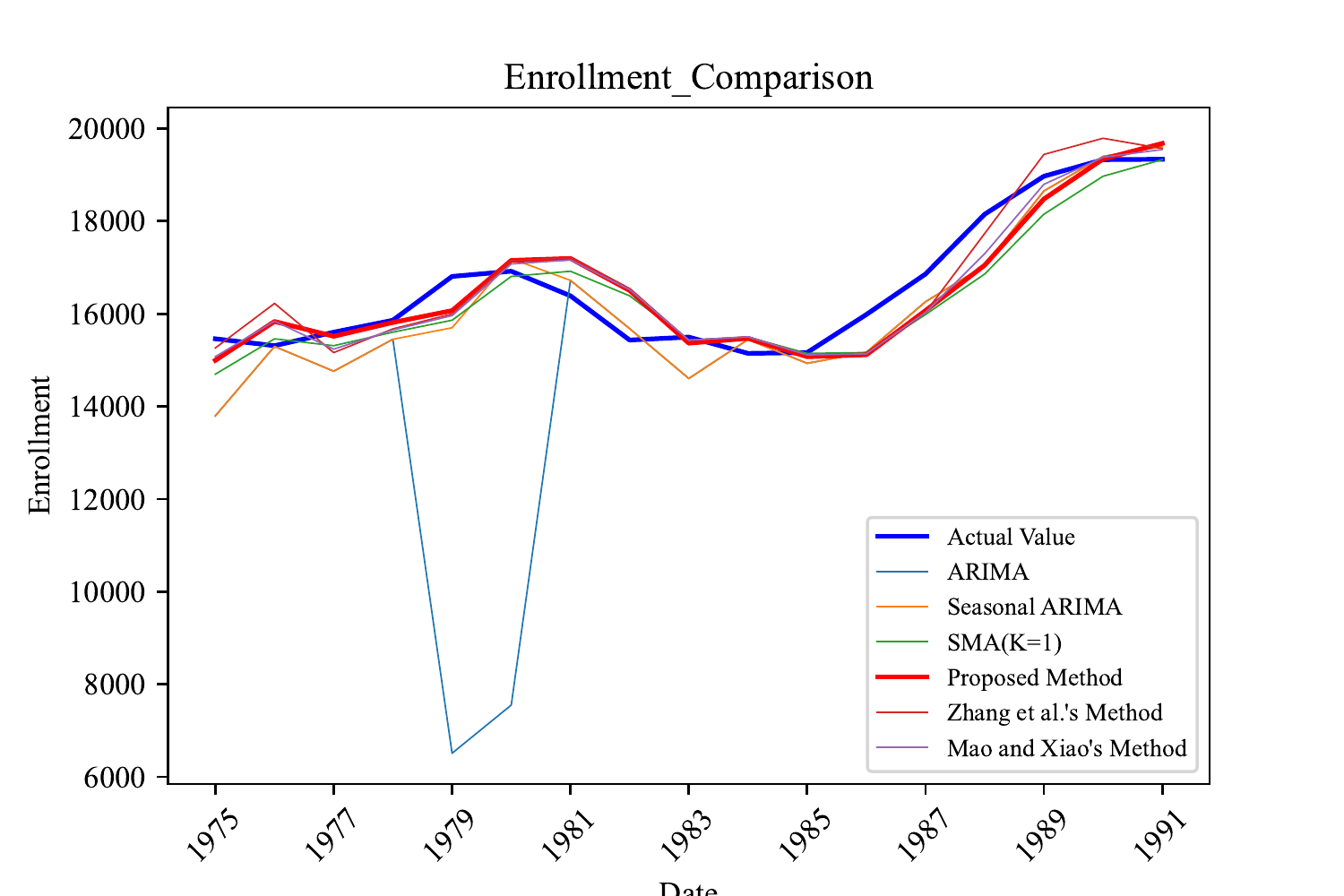}}
		\centerline{\includegraphics[scale=1]{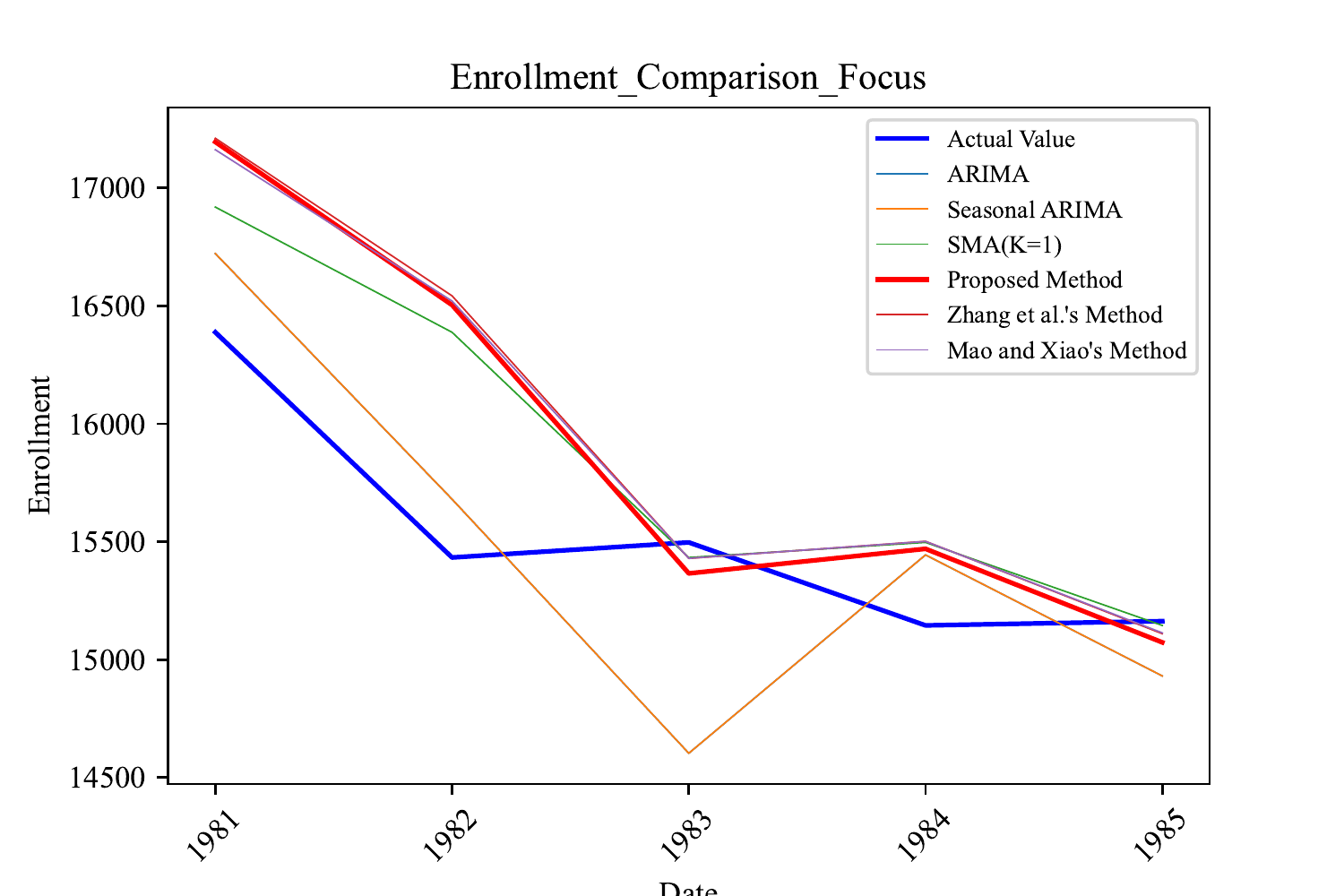}}
		\caption{The Enrollment and comparison of predicted values} 
	\end{figure}

	\begin{table}[htbp]
		\centering
		\setlength{\tabcolsep}{1mm}
		\begin{tabular}{cccccc}
			\hline
			Prediction Method & MAD       & MAPE   & SMAPE   & RMSE      & NRMSE     \\ \hline
			SMA(k=1)  \cite{guan2017two}         & 1611.1598 & 9.7146 & 12.4878 & 3353.3610 & 4922.3730 \\
			Zhang et al.  \cite{zhang2017novel}   & 498.2775  & 3.0513 & 3.0452  & 586.9486  & 861.5774  \\
			Mao and Xiao   \cite{mao2019time}             & 471.0208  & 2.9060 & 2.9199  & 574.4591  & 843.2442  \\
			ARIMA  \cite{tseng2002combining}            & 595.4259  & 3.6910 & 3.7956  & 743.3847  & 1091.2087 \\
			Seasonal ARIMA  \cite{tseng2002fuzzy}    & 524.6667  & 3.1828 & 3.2309  & 649.0047  & 952.6691  \\
			Proposed Method   & 490.4143  & 2.9943 & 3.0110  & 599.3861  & 879.8343  \\ \hline
		\end{tabular}
	\caption{The errors of the proposed method in the prediction of Enrollment}
	\end{table}

	\subsection{Analysis}
	In the large stock market, the methods of complex network  \cite{zhang2017novel,mao2019time} ARIMA  \cite{tseng2002combining} and seasonal ARIMA  \cite{tseng2002fuzzy} cannot be calculated, and the prediction effect of the proposed method is better than that of SMA  \cite{guan2017two}, and the prediction effect is better. In terms of logic time complexity, the complexity of the proposed method is only $O\left(1\right)$. At the same time, in terms of small data sets in other fields, the prediction effect of the proposed method is also better than that of complex network methods. The proposed method has both accuracy and performance.

	\section{Conclusion}
	In recent years, the fast prediction of big time series has received widespread attention. This article proposes a new time series forecasting method inspired by the evidence theory.The biggest feature of this method is that BPA is generated based on the fusion of time series information, and the predicted value is obtained through inverse calculation. At the same time, the calculation result can be inherited, which improves the calculation efficiency. With the high prediction performance,  big stock time series can be predicted. The proposed method has good predictive performance, and it can also be improved in many ways. Machine learning methods can be used to reduce the number of time points involved in calculations to improve performance again, and this work will continue to be optimized in the future.

	\section*{Acknowledgment}
	The authors greatly appreciate the reviewers' suggestions and the editor's encouragement. This research is supported by the National Natural Science Foundation of China (No.62003280).

	\bibliographystyle{elsarticle-num}
	\bibliography{References}

\end{document}